\documentclass{article}
\usepackage{comment}
\usepackage[preprint]{neurips_2026}
\usepackage[utf8]{inputenc}
\usepackage[T1]{fontenc}
\usepackage{hyperref}
\usepackage{url}
\usepackage{booktabs}
\usepackage{amsfonts}
\usepackage{amsmath}
\usepackage{amssymb}
\usepackage{nicefrac}
\usepackage{xcolor}
\usepackage{graphicx}
\usepackage{multirow}
\usepackage{array}
\usepackage{natbib}

\title{When Do LLMs Generate Realistic Social Networks? \\ A Multi-Dimensional Study of Culture, Language, Scale, and Method}

\author{
\textbf{Sai Hemanth Kilaru} \quad
\textbf{Sriram Theerdh Manikyala} \quad
\textbf{Raghav Upadhyay} \\
\textbf{Sri Sai Kumar Ramavath} \quad
\textbf{Srivika Nunavathu} \quad
\textbf{Dalal Alharthi} \\
University of Arizona. Tucson, AZ 85721 \\
\texttt{\{skilaru, manikyala, raghavupadhyay, srisairamavath45,} \\
\texttt{srivikanunavathu, dalharthi\}@arizona.edu}
}

\begin{document}

\maketitle

\begin{abstract}
Large language models (LLMs) are increasingly used as substitutes for human subjects in behavioral
simulations, including the generation of synthetic social networks. While prior work has shown that LLMs produce structurally realistic graphs, the extent to which their relational outputs depend on prompt design, cultural framing, prompt language, and model scale remains poorly characterized. Building on classical homophily theory and structural balance theory from network sociology, we formalize four LLM-based tie-formation mechanisms (sequential, global, local, iterative) as distinct conditional distributions over edge sets and quantify how each interacts with culture, language, and model scale. Holding a fixed roster of 50 demographically grounded personas constant, we generate 192 verified directed networks across four cultural contexts (United States,
India, Japan, Brazil), four prompt languages (English, Spanish, Hindi, Japanese), three GPT-4.1
model variants, and four prompting architectures, with two seeds per condition. We find that
cultural framing produces measurable, non-uniform shifts in inbreeding homophily and
largest-component connectivity; that political affiliation dominates tie formation under three of
four methods while the global method substitutes age, revealing prompt architecture as a
substantive sociological variable; that model scale produces a stable divergence ranking that
reproduces across all three studies, with the smallest variant behaving qualitatively rather than
merely noisily differently; and that prompt language alone, with culture and personas held fixed,
shifts religion homophily sharply (especially under Hindi prompting) while leaving political
homophily nearly invariant. LLM-generated networks match real social graphs on clustering and
modularity better than Erd\H{o}s\,/\,R\'enyi, Barab\'asi\,/\,Albert, or Watts\,/\,Strogatz baselines, yet
encode demographic biases that exceed empirically observed levels. The apparent neutrality of
LLM-based social simulation is illusory: prompt design choices that are often treated as
implementation details encode substantive sociological assumptions.
\end{abstract}

\section{Introduction}
\label{sec:intro}

The study of social networks, that is, how individuals form relationships, cluster into communities,
and propagate information, has long been a central concern across sociology, political science, and
computational social science. Generating realistic synthetic networks has traditionally required
either costly large-scale empirical surveys or mathematical generative models such as preferential
attachment \citep{barabasi1999emergence}, stochastic blockmodels \citep{holland1983stochastic}, or
exponential random graph models \citep{goodreau2009birds}. The former are expensive and
privacy-sensitive; the latter sacrifice the demographic and behavioral nuance that characterizes
real human social systems.

LLMs offer a third path. By encoding socio-cultural knowledge from human-generated text, they can
be prompted to simulate the fundamentally social act of choosing friends. The use of LLMs as
substitutes for human survey respondents and behavioral subjects has expanded rapidly in the past
two years \citep{argyle2023out,aher2023using,park2023generative,horton2023large}, but the
\emph{relational} dimension, namely what kinds of social structures emerge when LLM agents are
tasked with forming ties, has received much less systematic study. \citet{chang2024llms} have shown
that GPT-class models produce networks with realistic structural signatures while simultaneously
overestimating political homophily relative to empirical baselines. We extend this line of inquiry
along four orthogonal dimensions, each connected to a separate concern in the literature on LLM
behavior. We ask, first, whether \emph{cultural context} embedded in the prompt alters network
structure and homophily, with persona roster and prompt language held fixed; second, which
\emph{demographic dimension} (gender, age, race or ethnicity, religion, political affiliation, or
shared interests) most strongly drives tie formation, and whether this is stable across generation
methods; third, whether different \emph{LLM model sizes} produce interchangeable networks or
diverge in structured ways; and fourth, whether \emph{prompt language alone}, with culture and
personas fixed, shifts homophily and topology.

\paragraph{Contributions.}
First, we recast LLM-based network generation as a family of formally distinct conditional
distributions over edge sets, indexed by a prompt architecture that conditions on different
amounts of relational context. Second, we ground the resulting empirical patterns in two
established sociological frameworks: the baseline / inbreeding decomposition of homophily
\citep{mcpherson2001birds} and structural balance theory
\citep{heider1946attitudes,cartwright1956structural}. Third, we contribute 192 verified directed
networks over a fixed 50-persona roster, spanning four cultures, four languages, three model
scales, and four prompting strategies. Fourth, we document a stable model-divergence ranking that
reproduces across three independent experiments, indicating that small-tier models are
qualitatively rather than quantitatively different. Finally, we show that homophily dimensions
respond \emph{asymmetrically} to language and culture: religion is highly language-sensitive
(especially under Hindi prompting), while political homophily is remarkably language-invariant.

\section{Related Work}
\label{sec:related}

Four threads of prior work converge on the present study: the graph-theoretic toolkit for
measuring network structure, the sociological theory of homophily and structural balance, the use
of LLMs as proxies for human respondents, and the documentation of cultural and linguistic
variation in LLM outputs.

\paragraph{Network structure and graph generators.}
Classical generative models, including Erd\H{o}s\,/\,R\'enyi random graphs \citep{erdos1959random},
Watts\,/\,Strogatz small worlds \citep{watts1998collective}, and Barab\'asi\,/\,Albert preferential
attachment \citep{barabasi1999emergence}, capture stylized facts of degree distributions and path
lengths but ignore demographic structure. Stochastic blockmodels \citep{holland1983stochastic}
introduced demographic-aware mixing matrices, and modularity-based community detection
\citep{newman2006modularity} provides a complementary lens on community structure. Comprehensive
treatments of social and economic networks integrate these tools with sociological theory
\citep{jackson2008social,easley2010networks}, providing the disciplinary foundation against which
any new generative process, including LLM-based generation, should be evaluated.

\paragraph{Sociology of homophily and structural balance.}
\citet{mcpherson2001birds} systematized homophily and crucially distinguished \emph{baseline}
homophily (the share of same-group ties induced merely by group sizes) from \emph{inbreeding}
homophily (the residual preference beyond what composition predicts), building on
\citet{lazarsfeld1954friendship}. \citet{granovetter1973strength} established the structural
importance of weak ties for information diffusion across group boundaries;
\citet{kossinets2006empirical} provided one of the first large-scale longitudinal analyses of an
evolving network; \citet{moody2001race} documented persistent racial homophily in U.S.\ schools;
and \citet{currarini2009economic} formalized homophily as the equilibrium of a friendship-formation
game, distinguishing preference-driven from opportunity-driven sorting.
\citet{goodreau2009birds} used exponential random graph models to disentangle homophily from
triadic closure in adolescent networks. Structural balance theory
\citep{heider1946attitudes,cartwright1956structural} provides the complementary mechanism for the
triadic closure observed in real networks. Empirically, platform studies report elevated but
bounded political homophily online \citep{bakshy2015exposure,conover2011political}, in keeping with
the broader phenomenon of affective polarization \citep{iyengar2019origins}, against which our
findings can be calibrated.

\paragraph{LLMs as proxies for human respondents.}
\citet{argyle2023out} coined \emph{algorithmic fidelity} to describe GPT-3's reproduction of
demographic survey distributions; \citet{santurkar2023whose} and \citet{durmus2024towards} showed
that LLM opinion distributions align unevenly with demographic and national subgroups;
\citet{aher2023using} demonstrated through ``Turing Experiments'' that LLMs can replicate classic
human-subject studies in psychology, linguistics, and behavioral economics;
\citet{horton2023large} treated LLMs as simulated economic agents; \citet{park2023generative}
showed that LLM agents can drive believable interactive social simulations;
\citet{tornberg2023simulating} simulated social-media feeds with LLM-driven users; and
\citet{hewitt2024predicting} demonstrated predictive validity of LLM simulations for social-science
experimental treatment effects. The validity of these substitutions is contested:
\citet{bisbee2024synthetic} and \citet{dillion2023can} document systematic biases that limit LLMs'
usefulness as drop-in replacements for human participants. We take a relational rather than
individual view, asking whether the \emph{aggregate relational structure} produced by LLM agents
matches real-network statistics.

\paragraph{Cultural and linguistic variation in LLM outputs.}
LLMs trained predominantly on English text have been shown to reflect WEIRD
\citep{henrich2010weirdest} cultural assumptions \citep{johnson2022ghost}.
\citet{cao2023assessing}, \citet{naous2024beer}, and \citet{tao2024cultural} document systematic
cross-cultural misalignment between LLM outputs and the values of non-Western societies, while
\citet{hofmann2024dialect} shows that decisions made by language models can be covertly racist
when conditioned on dialect. The broader context is provided by foundation-model risk surveys
\citep{bender2021dangers} and by documentation of persistent group-targeted biases such as
anti-Muslim association \citep{abid2021persistent}. These findings motivate our isolation of
prompt language as a single experimental variable. Methodologically, our characterization of the
global method is informed by results on long-context attention degradation \citep{liu2023lost} and
on prompt-architecture effects \citep{brown2020language,wei2022chain}. Our most direct foundation
is \citet{chang2024llms}, whose four prompting methods we adopt; we extend that work by treating
cultural context, prompt language, model size, and prompting strategy as independent experimental
variables.

\section{Theoretical Framework and Methodology}
\label{sec:method}

The four prompting architectures we study are not merely engineering variants of a single
generation procedure; they correspond to distinct social-cognitive scenarios with distinct
sociological pedigrees. Sequential generation places the LLM in the role of an ego responding to
a roster, a mode familiar from name-generator surveys; global generation places it in the role of
an omniscient observer assigning ties; local generation enforces bounded context, mirroring the
focused-organization account of social ties whereby individuals sort within shared activity foci
\citep{currarini2009economic}; and iterative generation mirrors the dyadic update dynamics
implicit in network-evolution models \citep{kossinets2006empirical,goodreau2009birds}. The four
architectures therefore induce distinct conditional distributions over edge sets that reflect
different assumptions about how social cognition is decomposed. We define each formally below,
then describe the homophily metrics, experimental design, and verification protocol.

\subsection{Formal Setup}

Let $\mathcal{V} = \{v_1, \ldots, v_n\}$ denote a fixed roster of $n=50$ personas. Each persona is
a structured profile $v_i \in \mathcal{A} = \mathcal{A}_g \times \mathcal{A}_a \times \mathcal{A}_r
\times \mathcal{A}_R \times \mathcal{A}_p \times \mathcal{A}_I$, with attribute spaces for gender,
age, race or ethnicity, religion, political affiliation, and a multi-set of interests. Let
$f_\theta: \Sigma^{*} \to \Delta(\Sigma^{*})$ denote the LLM, parameterized by model identity
$\theta \in \{\mathrm{nano},\mathrm{mini},\mathrm{full}\}$ and operated at fixed sampling temperature
$\tau = 0.8$. A generated social network is a directed graph $G = (\mathcal{V}, E)$, with
$E \subseteq \mathcal{V} \times \mathcal{V}$ specifying directed friendship nominations.

We formalize the four prompting architectures as four families of conditional distributions over
edge sets, induced by prompt templates $\pi_M$ for $M \in \{\mathrm{seq}, \mathrm{glob},
\mathrm{loc}, \mathrm{iter}\}$. The methods differ in how much relational context they condition
on and in whether tie decisions factorize across source nodes.

\emph{Sequential generation} prompts each persona independently with the rest of the roster, so
the model returns $S_i \sim f_\theta(\pi_{\mathrm{seq}}(v_i, \mathcal{V} \setminus \{v_i\}))$ and
the joint distribution \emph{factorizes} by source:
\begin{equation}
\Pr_{\mathrm{seq}}\!\left(e_{ij}=1 \mid \mathcal{V}\right)
= \mathbb{E}_{S_i}\!\bigl[\mathbf{1}\{v_j \in S_i\}\bigr],
\qquad
\Pr_{\mathrm{seq}}(E \mid \mathcal{V}) = \prod_{i=1}^{n} \Pr(S_i \mid v_i, \mathcal{V}).
\label{eq:seq}
\end{equation}
\emph{Global generation} presents the full roster in one prompt and samples the entire edge set
jointly,
\begin{equation}
\Pr_{\mathrm{glob}}(E \mid \mathcal{V}) = f_\theta(\pi_{\mathrm{glob}}(\mathcal{V}))[E],
\label{eq:glob}
\end{equation}
which does not factorize: the joint sample reflects whatever attentional structure $f_\theta$
imposes on the long context \citep{liu2023lost}. \emph{Local generation} restricts each persona's
choices to a curated neighborhood $\nu(v_i) \subset \mathcal{V} \setminus \{v_i\}$ of size $k$, so
$\Pr_{\mathrm{loc}}(e_{ij}=1) = \mathbf{1}\{v_j \in \nu(v_i)\} \cdot
\mathbb{E}_{S_i \sim f_\theta(\pi_{\mathrm{loc}}(v_i, \nu(v_i)))}\!\bigl[\mathbf{1}\{v_j \in S_i\}\bigr]$,
mirroring bounded sociological decision contexts. Finally, \emph{iterative generation} defines a
Markov chain over edge sets,
\begin{equation}
E^{(t)} \sim f_\theta\!\left(\pi_{\mathrm{iter}}(\mathcal{V}, E^{(t-1)})\right),
\qquad t = 1, \ldots, T,
\label{eq:iter}
\end{equation}
whose stationary distribution depends jointly on $f_\theta$ and $\pi_{\mathrm{iter}}$, and whose
factorization properties depend on the prompt's per-round encoding of $E^{(t-1)}$.

These methods differ along two structural axes: whether the joint distribution factorizes by
source node (yes for sequential and local, no for global and iterative) and whether each tie
decision conditions on the global persona context. A non-factorizing method permits, in
principle, the kind of triadic-balance reasoning predicted by \citet{cartwright1956structural},
but the long context that enables it also exposes the model to attentional degradation that can
shift which attributes drive tie formation.

\subsection{Homophily Measurement}

For attribute $a \in \{g, a, r, R, p, I\}$, partition the roster as $\mathcal{V} = \bigsqcup_g
\mathcal{V}^{(a)}_g$, where $\mathcal{V}^{(a)}_g$ collects personas with attribute value $g$. The
random baseline probability that a directed pair shares value $g$ on attribute $a$, accounting for
self-exclusion, is
\begin{equation}
p^{\mathrm{base}}_a = \sum_{g} \frac{|\mathcal{V}^{(a)}_g|}{n} \cdot
\frac{|\mathcal{V}^{(a)}_g| - 1}{n - 1}.
\label{eq:baseline}
\end{equation}
Following the baseline / inbreeding decomposition of \citet{mcpherson2001birds}, we measure
\emph{inbreeding homophily} as the same-group ratio
\begin{equation}
H_a(G) = \frac{1}{|E|\,p^{\mathrm{base}}_a}
\sum_{(i,j) \in E} \mathbf{1}\{a(v_i) = a(v_j)\}.
\label{eq:samegroup}
\end{equation}
Values $H_a > 1$ indicate same-group edge frequencies in excess of what roster composition alone
would predict; values near 1 indicate purely baseline homophily. This separation is essential for
interpreting cross-condition shifts: a change in $H_a$ across cultural framings, with the roster
held constant, isolates the LLM's tendency to amplify or suppress dimension $a$ as a tie-formation
cue, distinct from the demographic composition of the population.

\subsection{Experimental Design and Metrics}

The study is organized as three controlled experiments, each holding the persona roster fixed and
varying one independent variable at a time. The cultural study varies cultural framing across four
settings while keeping language at English and method at sequential; the method study introduces
the global, local, and iterative architectures across the same four cultures; and the language
study fixes culture to the United States and varies prompt language across English, Spanish,
Hindi, and Japanese under all four methods. Each condition is run with two seeds across three
GPT-4.1 model variants (\texttt{gpt-4.1-nano}, \texttt{gpt-4.1-mini}, \texttt{gpt-4.1}) at fixed
sampling temperature $\tau = 0.8$. The full design produced 192 verified networks
(Table~\ref{tab:matrix}). The persona roster, sampled once from U.S.\ adult demographic marginals
\citep{uscensus2023}, and the prompt templates for each method are described in
Appendices~\ref{app:roster}\,/\,\ref{app:prompts}.

We measure topology using density, average clustering coefficient, the proportion of nodes in the
largest connected component (denoted $\mathrm{lcc}$), average shortest path length, and modularity
\citep{newman2006modularity}, and homophily by the same-group ratio in
Eq.~\eqref{eq:samegroup}. For cross-condition comparison we use \emph{pairwise edge distance}, the
proportion of possible edges on which two networks disagree, and the
Kolmogorov\,/\,Smirnov statistic for degree distributions.

\begin{table}[t]
\caption{Experimental matrix. Each cell entry counts verified directed networks of 50 nodes.
Holding the roster fixed makes any structural difference attributable to the manipulated variable.}
\label{tab:matrix}
\centering
\small
\begin{tabular}{lcccc}
\toprule
Study & Cultures & Languages & Methods & Networks \\
\midrule
Cultural & 4 (US, India, Japan, Brazil) & 1 (English) & 1 (sequential) & 24 \\
Method   & 4 & 1 (English) & 3 (global, local, iterative) & 72 \\
Language & 1 (US, fixed) & 4 (En, Es, Hi, Ja) & 4 (all) & 96 \\
\midrule
Total    & & & & 192 \\
\bottomrule
\end{tabular}
\end{table}

\section{Results}
\label{sec:results}

\subsection{RQ1: Cultural Framing Alters Network Structure and Homophily}

Holding language fixed at English while varying cultural context produces measurable structural
shifts under all four generation methods. Political affiliation is the most culture-sensitive
homophily dimension under sequential, local, and iterative generation, with the dominant predictor
of tie formation reordering across cultures. Among topology metrics, $\mathrm{lcc}$, the proportion
of nodes in the largest connected component, is the metric most sensitive to cultural framing,
indicating that cultural context most strongly affects the global \emph{cohesiveness} of the
simulated society rather than simply its edge count. Sequential and local methods produce the most
cohesive networks (typical $\mathrm{lcc}$ between 0.75 and 0.92), while global and iterative
methods produce more fragmented networks that rarely exceed 0.75. India yields the highest
connectivity under sequential generation, approaching $\mathrm{lcc} \approx 0.92$, while Japan
produces the lowest local-method value across all culture and method combinations.

\begin{figure}[t]
\centering
\includegraphics[width=0.49\linewidth]{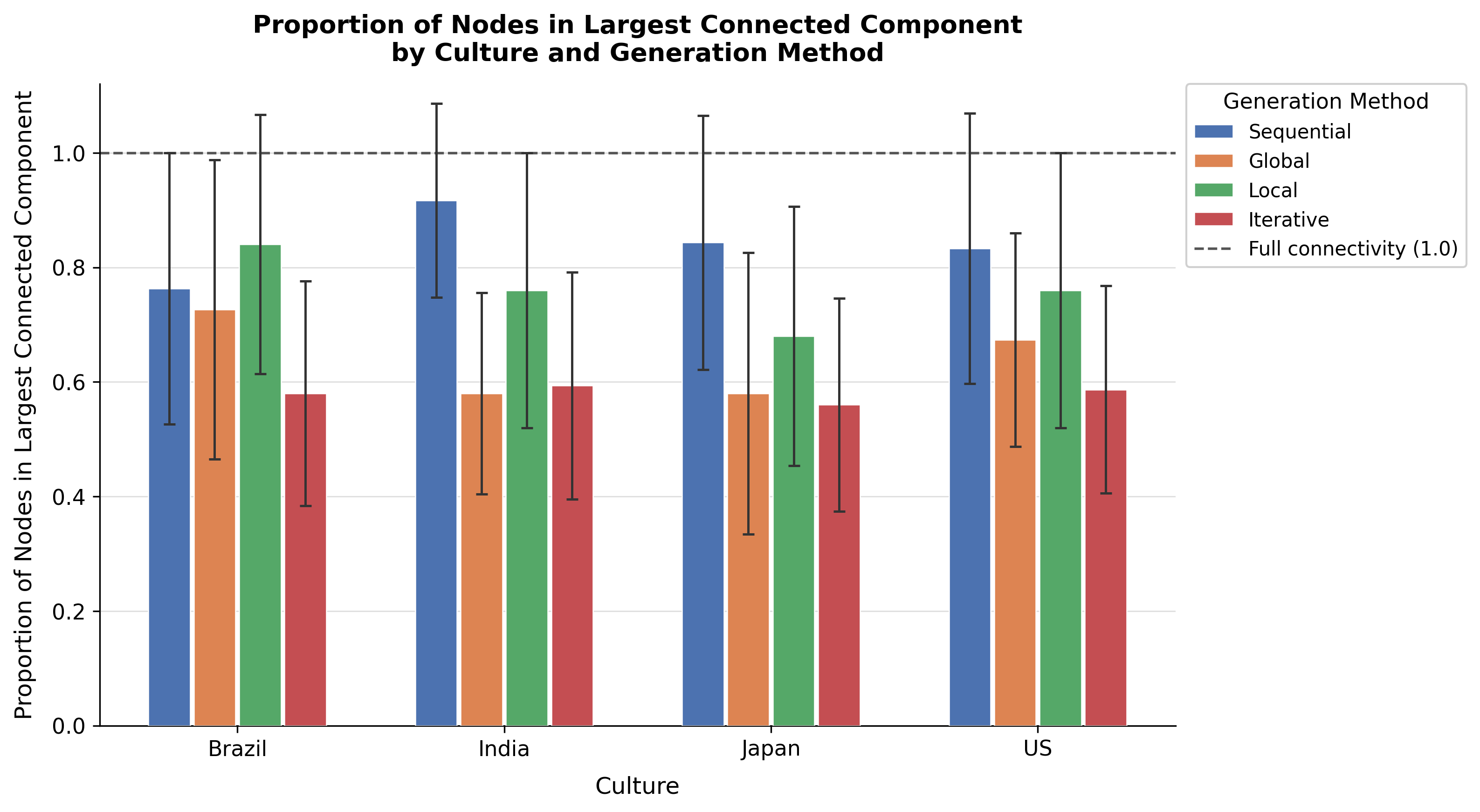}\hfill
\includegraphics[width=0.49\linewidth]{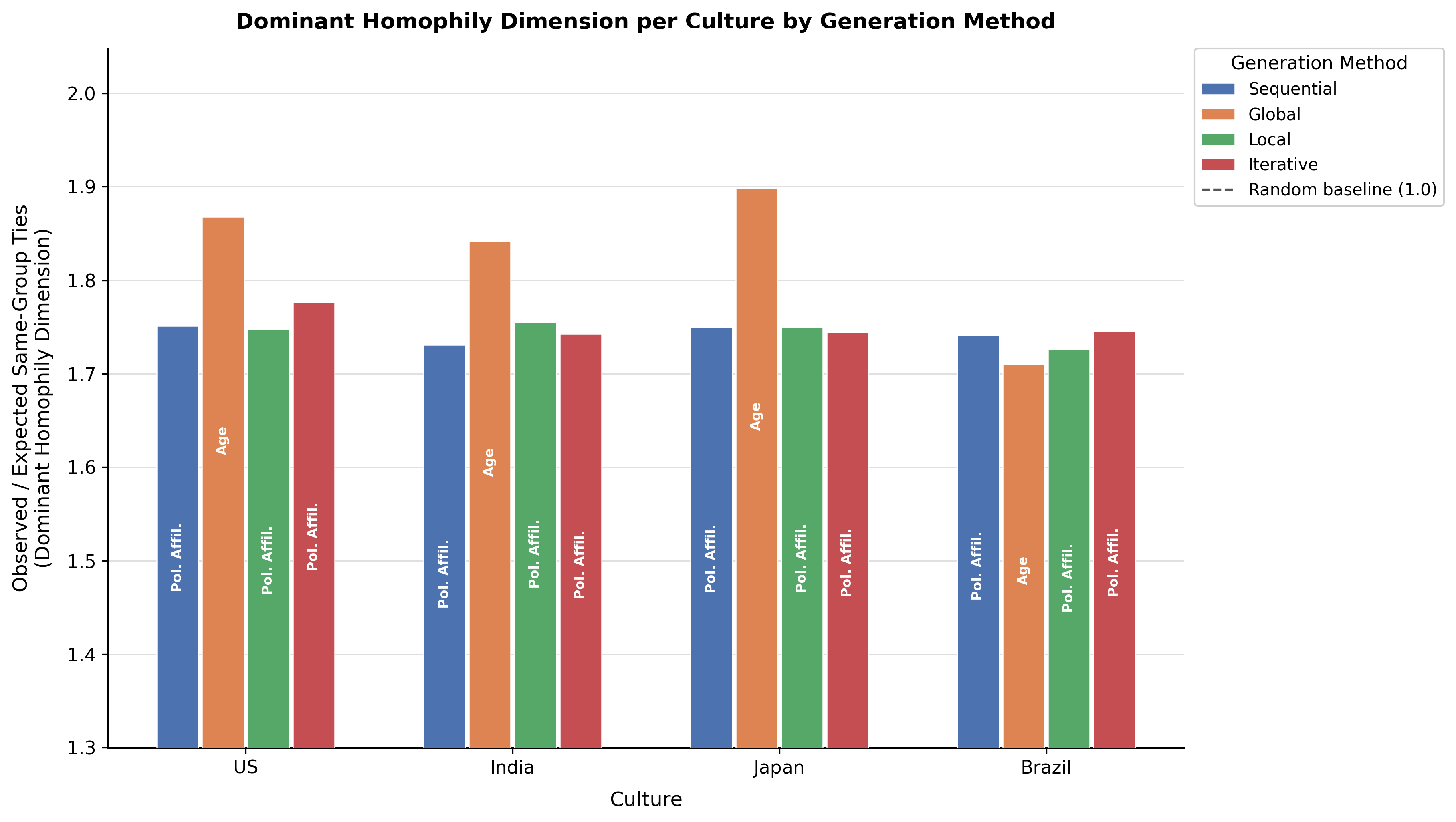}\\[0.3em]
\includegraphics[width=0.78\linewidth]{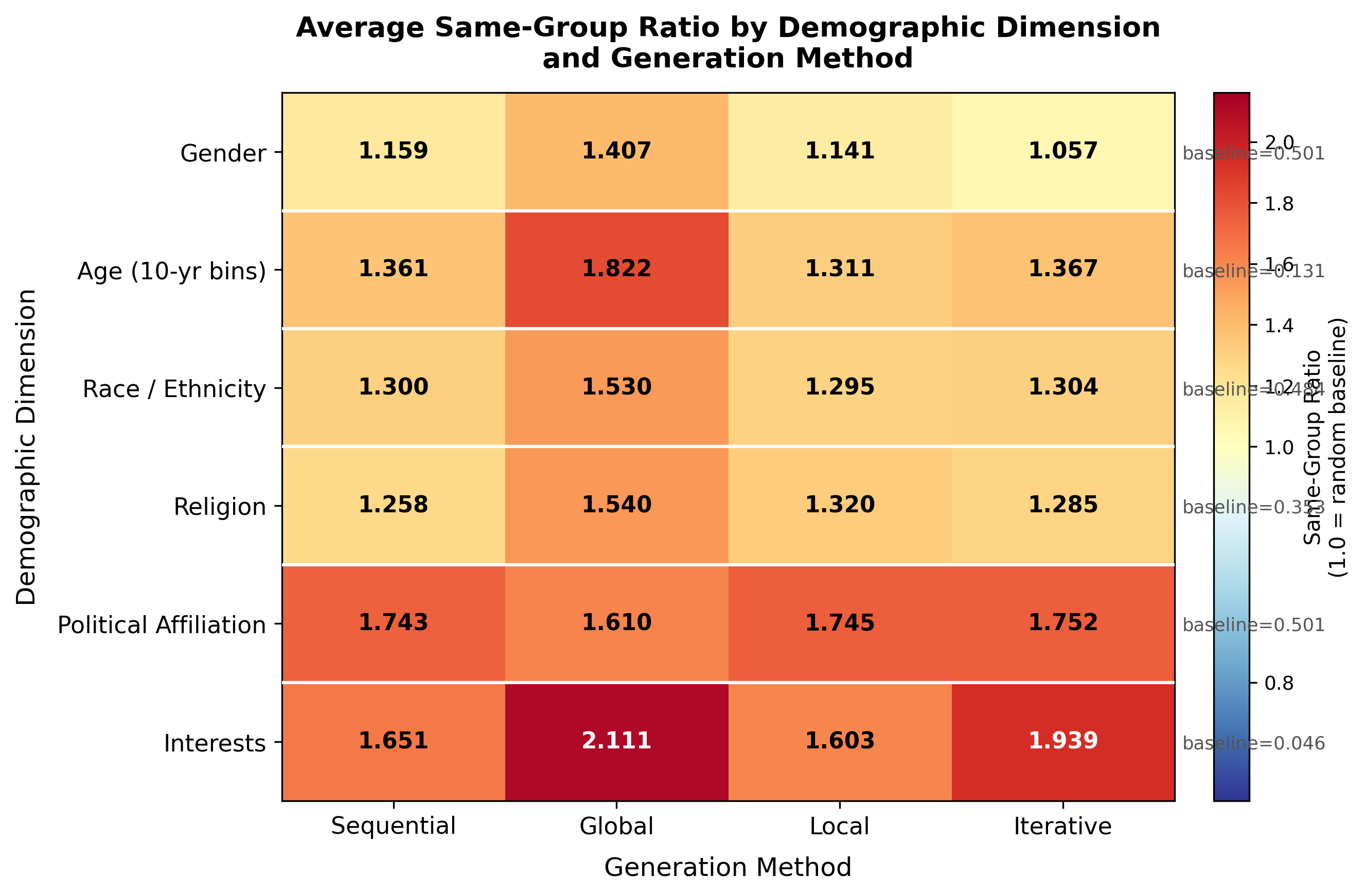}
\caption{Topology and homophily across the culture-by-method design (GPT-4.1-mini). Top left:
proportion of nodes in the largest connected component by culture and method, with $\pm 1$\,SE
across seeds. Top right: dominant homophily dimension per culture by method; every bar exceeds the
random baseline of 1.0, with political affiliation dominating under sequential, local, and
iterative and the global method substituting age in three of four cultures. Bottom: average
same-group ratio by demographic dimension and method, where the global method is the systematic
outlier, elevating age (1.822) and interests (2.111) while depressing political affiliation
(1.610).}
\label{fig:rq1-rq2}
\end{figure}

The cultural-framing effect is not an artifact of any single prompting strategy: it appears across
all four methods. Its magnitude, however, varies. The global method, which is the sparsest overall
(mean density $\approx 0.056$), shows the most muted cultural differentiation, while the iterative
method (mean density $\approx 0.182$) shows the sharpest cultural shift in $\mathrm{lcc}$.
Figure~\ref{fig:rq1-rq2} (top panels) summarizes the topology and dominant-homophily structure
across the culture-by-method design.

\subsection{RQ2: Political Affiliation Dominates, except under Global Generation}

Across the sequential, local, and iterative methods, political affiliation produces the highest
same-group ratio in the majority of culture, method, and model combinations. The pattern persists
across all four cultural contexts under English prompting and across all four languages under
fixed-United-States framing, exceeding the random baseline of Eq.~\eqref{eq:baseline} in virtually
every condition. The consistency of this finding across three distinct generation architectures
and three model sizes provides strong evidence that LLM political-homophily inflation is not a
methodological artifact of any single prompting strategy. The magnitudes are consistent with, but
exceed, the empirically observed levels of online political homophily reported by
\citet{bakshy2015exposure} and \citet{conover2011political}.

The global method is the systematic exception. Rather than elevating political affiliation, global
generation most often produces \emph{age} as the dominant homophily dimension. Two complementary
mechanisms can be invoked. First, when all 50 personas are presented in a single context, the
model's attention is plausibly drawn to visually salient surface features (age is parsed easily
when scanning a long list), while the subtler political-alignment signal that emerges from focused
pairwise reasoning is diluted; this is consistent with documented long-context attention
degradation \citep{liu2023lost}. Second, the joint factorization in Eq.~\eqref{eq:glob} permits
triadic reasoning of the kind described by structural balance theory
\citep{cartwright1956structural}, but does so over whichever cue the model treats as most salient
in the long context, which empirically is age rather than party.

The bottom panel of Figure~\ref{fig:rq1-rq2} confirms that all six demographic dimensions exhibit
above-chance homophily under all four methods. The global method exhibits both the highest age
ratio (1.822) and the lowest political-affiliation ratio (1.610) among methods, and it produces
the highest interest ratio (2.111). Gender is the weakest homophily signal across the board,
falling as low as 1.057 under iterative generation. We note that raw same-group ratios across
dimensions are not directly comparable without accounting for each dimension's roster baseline:
interests in particular benefits from a low chance baseline ($p^{\mathrm{base}}_I = 0.046$), which
amplifies its apparent ratio.

\subsection{RQ3: Models Are Not Interchangeable, and Their Divergence Is Stable}

\begin{figure}[t]
\centering
\includegraphics[width=0.45\linewidth]{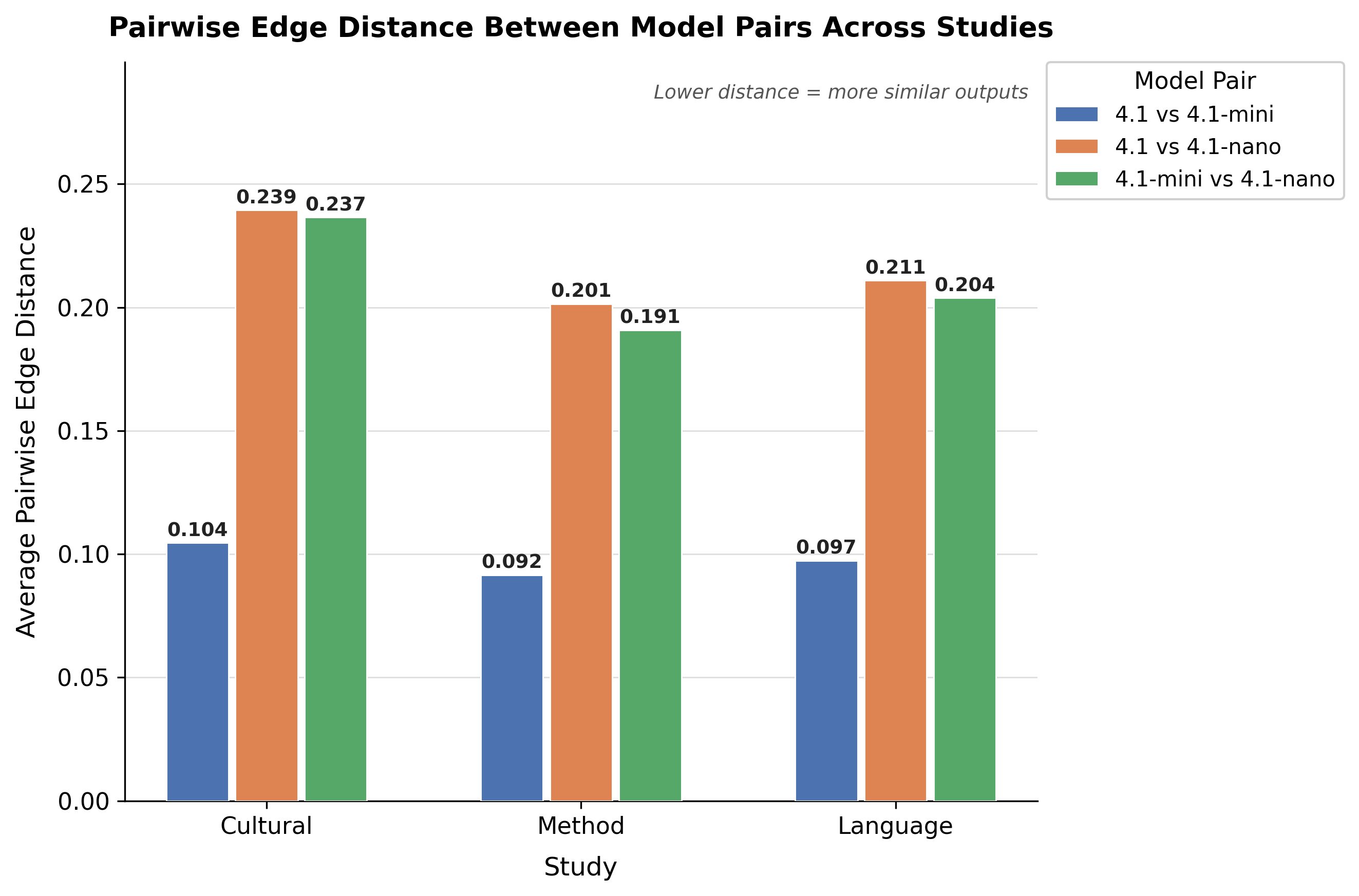}\hfill
\includegraphics[width=0.53\linewidth]{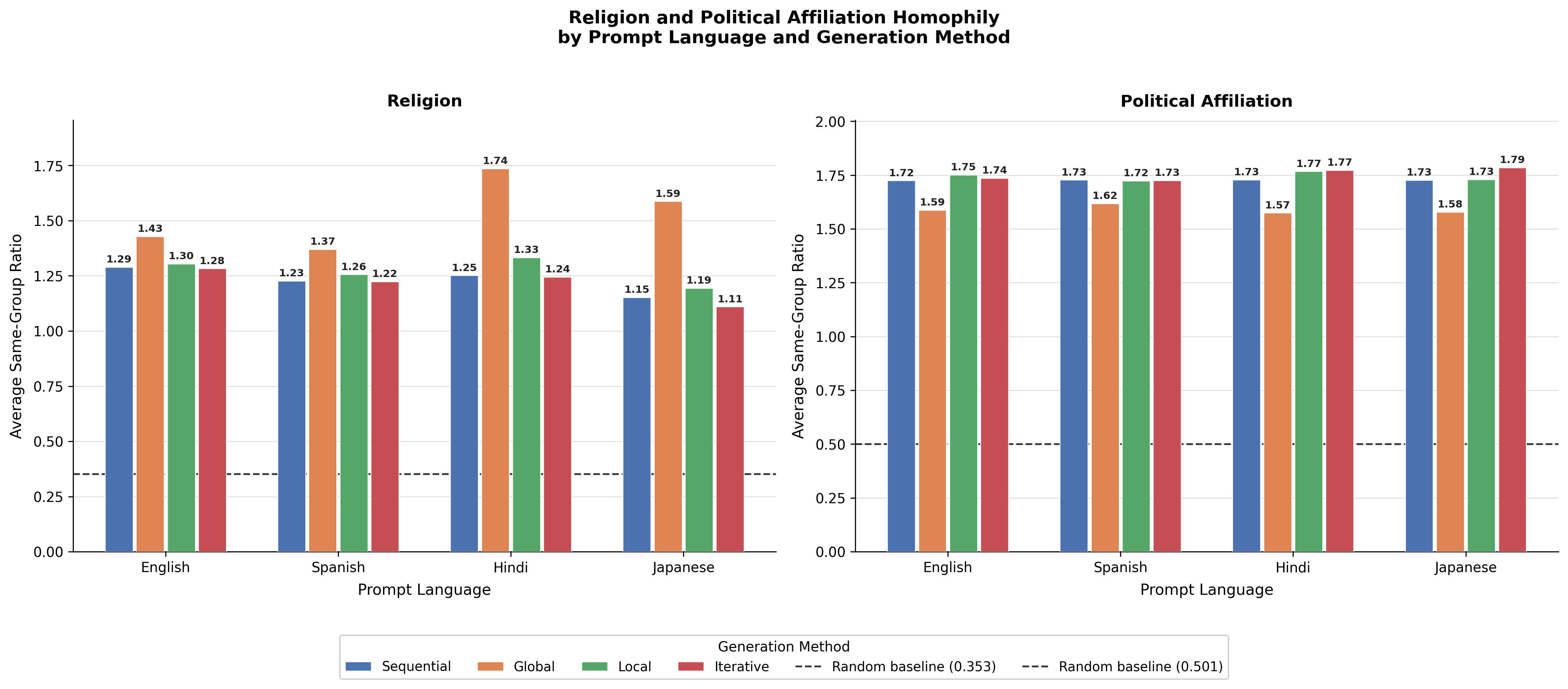}
\caption{Left: pairwise edge distance between model pairs across the three studies. The
\texttt{4.1}\,/\,\texttt{mini} pair is consistently the closest; both pairs involving
\texttt{nano} cluster together far above it, indicating that \texttt{nano} is a structurally
distinct generator rather than a noisy version of the larger models. Right: religion and political
affiliation homophily by prompt language and generation method. Religion shifts dramatically with
language (peaking at 1.74 under Hindi prompting with the global method), while political
affiliation remains in the narrow band 1.57 to 1.79 across all four languages.}
\label{fig:rq3-rq4}
\end{figure}

Pairwise edge distance reveals a strikingly stable ranking that reproduces across all three studies
without exception:
\begin{equation}
d(\mathrm{full}, \mathrm{mini}) \;<\; d(\mathrm{mini}, \mathrm{nano}) \;<\; d(\mathrm{full}, \mathrm{nano}).
\label{eq:edgedist}
\end{equation}
The closest pair (\texttt{gpt-4.1} versus \texttt{gpt-4.1-mini}) sits at $d \in [0.09, 0.10]$
across studies, while the two pairs involving \texttt{gpt-4.1-nano} cluster together near
$d \in [0.19, 0.24]$. The gap between the closest and farthest pair, roughly $0.11$ to $0.14$
across studies, exceeds seed-to-seed noise within a model by a substantial margin
(Figure~\ref{fig:rq3-rq4}, left). The practical implication is that substituting
\texttt{gpt-4.1-nano} for \texttt{gpt-4.1}, for cost reasons, does not produce a noisier version of
the same network; it produces a meaningfully \emph{different} network. By contrast,
\texttt{gpt-4.1} and \texttt{gpt-4.1-mini} are close enough that mini may serve as a reasonable
cost / performance substitute under the conditions tested here.

While edge-level agreement varies, all three models agree on the qualitative homophily ranking: political affiliation dominates under sequential, local, and iterative generation, while age dominates under global generation. The disagreement is in the fine-grained relational structure, not in the broad demographic character of the generated networks.
The stable divergence ranking of Eq.~\eqref{eq:edgedist} indicates that relational
competence does not improve smoothly with model scale in the way perplexity-based
metrics do: the smallest variant produces a categorically different network rather
than a noisier version of what larger models produce, complicating the assumption
that scaling alone resolves the demographic distortions we document.

\subsection{RQ4: Prompt Language Shifts Religion, Not Politics}

With cultural framing fixed to the United States and the persona roster held constant, varying
prompt language across English, Spanish, Hindi, and Japanese reveals a distinct asymmetry. The
largest language-driven homophily shift appears in the \emph{religion} dimension, a notably
different pattern from RQ1, where political affiliation was the most culture-sensitive dimension.
Hindi prompting elevates religion homophily substantially relative to the English baseline, and
under the global method the religion same-group ratio reaches 1.74 in Hindi. Japanese prompts
produce a distinct profile in which age becomes more prominent, mirroring the global method's
behavior and suggesting that certain linguistic framings activate similar attentional shifts in
the model as certain prompting architectures.

Political affiliation is remarkably language-invariant, with values ranging narrowly between 1.57
and 1.79 across all four languages and four methods (Figure~\ref{fig:rq3-rq4}, right). The closest
topology pair is Hindi and Japanese, while the most distant pair is Japanese and Spanish, an
asymmetry that does not map cleanly onto any obvious linguistic distance metric and that suggests
the model's social behavior under different languages reflects training-data co-occurrence
patterns rather than surface linguistic similarity \citep{durmus2024towards,tao2024cultural}.
English and Spanish produce the most similar homophily profiles, consistent with their heavy joint
representation in GPT pre-training corpora. The global method again behaves distinctively,
producing the largest language-driven swings in religion homophily.

\subsection{Structural Realism: LLM-Generated Networks versus Classical Baselines}

\begin{figure}[t]
\centering
\includegraphics[width=0.9\linewidth]{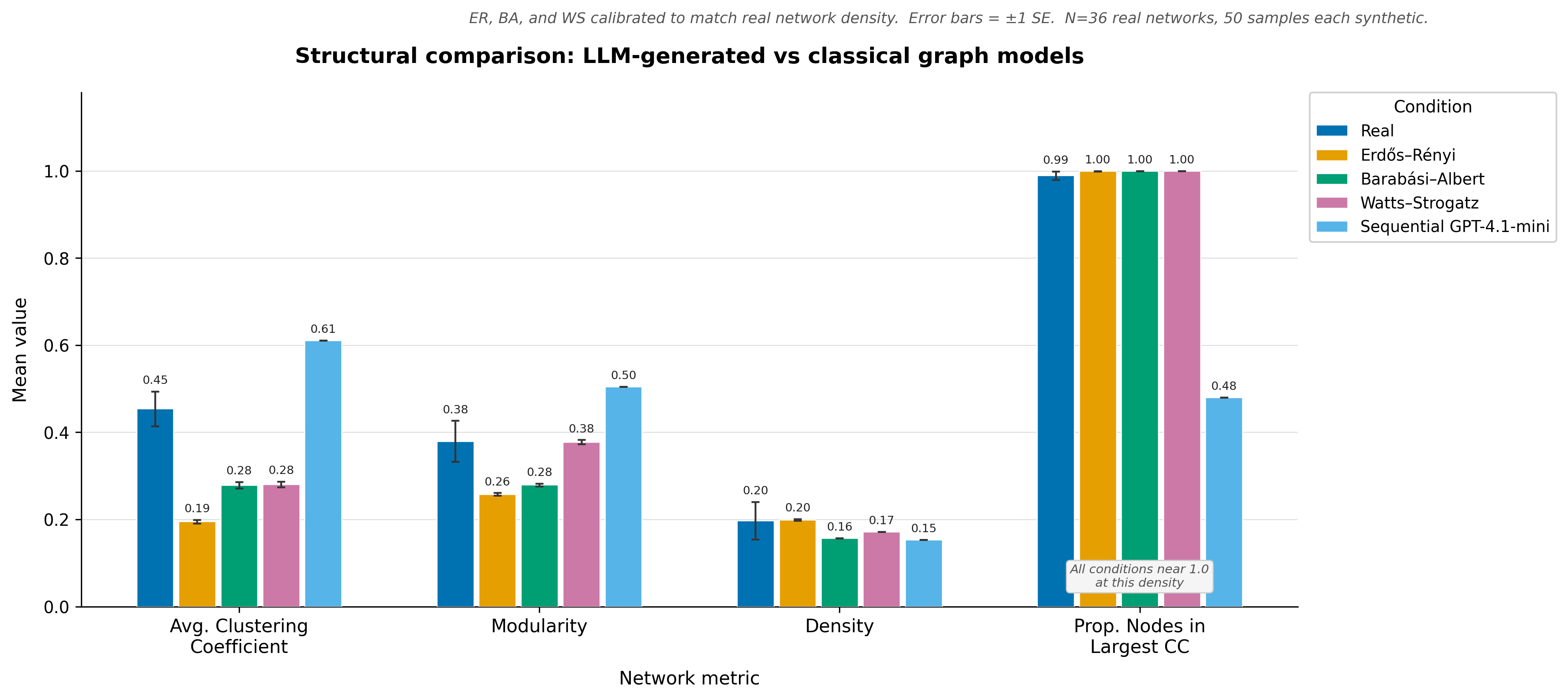}
\caption{Structural comparison: LLM-generated networks (Sequential GPT-4.1-mini, light blue) versus
real social networks (dark blue) and three classical generators calibrated to match real-network
density. LLM-generated networks match or exceed real networks on clustering (0.61 versus 0.45) and
modularity (0.50 versus 0.38); all three classical generators fall short on both. The
clustering-coefficient gap is most diagnostic: ER produces only 0.19 despite matching density,
confirming that random edge assignment cannot replicate the triadic-closure patterns shared by
real and LLM-generated networks.}
\label{fig:realism}
\end{figure}

To establish whether the homophily distortions documented above occur within structurally
plausible graphs, we benchmark LLM-generated networks against three classical generators fitted to
match real-network density: Erd\H{o}s\,/\,R\'enyi (ER) \citep{erdos1959random}, Barab\'asi\,/\,Albert (BA)
\citep{barabasi1999emergence}, and Watts\,/\,Strogatz (WS) \citep{watts1998collective}. Each was run
for 30 seeds. Across density, clustering, modularity, average shortest path, and $\mathrm{lcc}$,
LLM-generated networks more closely match real social networks than any classical baseline. The
most diagnostic gap is on the clustering coefficient: an ER graph at the empirical density produces
a clustering coefficient near 0.19, whereas LLM-generated networks reach 0.61, against 0.45 for
real networks; LLM-generated modularity (0.50) likewise exceeds both classical baselines and the
real benchmark (0.38) (Figure~\ref{fig:realism}). The clustering and modularity elevation is
precisely what structural balance theory \citep{heider1946attitudes,cartwright1956structural}
predicts when triadic closure pressure is strong, suggesting that LLMs internalize a
balance-theoretic prior even when not explicitly prompted to do so.

This is the central interpretive context for our homophily findings. LLM-generated networks are
not structurally implausible; they are, by graph-theoretic measures, \emph{more} realistic than
anything a classical model produces, while simultaneously encoding demographic biases that exceed
empirically observed levels \citep{bakshy2015exposure,iyengar2019origins}. The risk for downstream
users is therefore not noisy data; it is plausible-looking data with embedded social biases that
may pass surface-level inspection.

\section{Discussion}
\label{sec:discussion}
A unifying observation runs across our four results: each methodological lever pushes a
distinct aspect of the generated network. Cultural framing shifts cohesiveness; prompt
architecture shifts which demographic dimension dominates tie formation; model scale
shifts fine-grained edge structure; and prompt language shifts religion specifically. The
four are non-redundant, and fixing only one silently inherits the others' priors.

The four prompt architectures formalized in Eqs.~\eqref{eq:seq}--\eqref{eq:iter} induce
structurally distinct conditional distributions over edge sets, and their empirical signatures
match the theoretical distinctions. Sequential and local generation factorize by source and
produce the cohesive, political-homophily-dominated regime that prior work \citep{chang2024llms}
took as characteristic of LLM-generated networks; global generation, which does not factorize,
exposes the model to long contexts in which attentional dynamics \citep{liu2023lost} elevate
surface features (age, interests) over deeper relational cues; iterative generation, the Markov
chain of Eq.~\eqref{eq:iter}, tracks sequential at the homophily level but produces sparser
topologies. Read against \citet{mcpherson2001birds}, the consistent inflation of $H_a > 1$ across
every method and demographic dimension indicates genuine \emph{inbreeding} homophily rather than
the baseline component induced by roster composition; the magnitude is also consistent with the
preference-driven sorting predicted by economic models of friendship formation
\citep{currarini2009economic}, while exceeding the bounds estimated in empirical platform studies
\citep{bakshy2015exposure}. Read against \citet{cartwright1956structural}, the elevated clustering
and modularity relative to ER, BA, and WS baselines suggests that the LLM's representation of
friendship is implicitly balance-theoretic: triads tend toward closure not because the prompt
enforces it but because closure is a statistical regularity in the human-generated training text.
Cultural and linguistic framings then act as soft selectors over which attribute the balance
pressure operates upon: Hindi prompting elevates religion homophily because religion is a salient
axis of social organization in Hindi-language data; Japanese prompting elevates age homophily
because age stratification is similarly salient in Japanese discourse. Political homophily is
uniquely language-invariant, plausibly because political affiliation is encoded as a strong
relational feature throughout the English-dominated corpus and persists under translation rather
than being supplanted by language-specific salience. The downstream implication is that
researchers selecting among generation methods are not making a computational tradeoff but
choosing which demographic dimensions their networks will most strongly reflect; prompt
architecture, prompt language, and model scale should be reported as primary methodological
variables in LLM-based simulation, on equal footing with the persona roster.

\paragraph{Implications for computational social science.}
A growing literature treats LLMs as drop-in stand-ins for human survey respondents
\citep{argyle2023out,aher2023using,horton2023large,hewitt2024predicting}, but recent skeptical
analyses \citep{bisbee2024synthetic,dillion2023can} caution that LLM responses systematically
diverge from human distributions in ways that undermine straightforward substitution. Our results
contribute a relational complement to that debate. Even when individual-level outputs look
demographically plausible, the \emph{aggregate} edge structure they produce embeds sociologically
meaningful distortions whose magnitude and direction depend on prompt architecture, prompt
language, and model scale. Two practical recommendations follow. First, structural diagnostics,
including the same-group ratio decomposition of Eq.~\eqref{eq:samegroup}, clustering, and
modularity \citep{newman2006modularity}, should accompany any claim that an LLM-generated network
is a valid substitute for an empirical one. Second, sensitivity analyses across at least two prompt architectures and two prompt languages should be reported as a default, since fixing only one of these variables silently bakes in the demographic prior associated with that choice. Just as effect heterogeneity across demographic subgroups is now a standard component of human-subject research, heterogeneity across prompt-architecture and prompt-language conditions deserves a comparable place in LLM-based simulation. Treating these conditions as variables to report, rather than as defaults to fix, brings LLM-driven network generation into closer methodological alignment with the experimental and survey traditions on which it draws.
These concerns are not academic. LLM-agent simulations are increasingly used in
policy-relevant computational social science, and when LLM-generated networks
themselves become training or evaluation data for downstream models, the inflated
homophily we document closes a feedback loop in which synthetic biases self-reinforce.

\section{Limitations}
\label{sec:limitations}

The 50-persona roster, while grounded in U.S.\ census marginals, gives high-variance estimates for
minority categories and, since the same roster is reused across every cultural condition, limits
ecological validity for cross-cultural comparisons (though preserving their internal validity as
controlled tests of framing effects). The roster's near-balanced partisan composition (n{=}26 Republican, n{=}24 Democrat) further
leaves open whether the political-homophily inflation we report is amplified, attenuated,
or qualitatively altered under rosters with more skewed group sizes; this is a controlled question our single-roster design cannot resolve, and a natural extension of this work. All networks are generated rather than observed; realism is
established by distributional similarity to empirical benchmarks rather than against a ground-truth network among these personas, which does not exist. All experiments use the GPT-4.1
family at $\tau=0.8$, and two seeds per condition give an indication but not a high-precision
estimate of within-condition variance, particularly for global and iterative generation. Recent
work has cautioned that LLM stand-ins for human respondents systematically diverge from human
distributions \citep{bisbee2024synthetic,dillion2023can}; while our focus on relational rather
than individual-level statistics partially insulates us from those concerns, it does not
eliminate them. Cross-family extension, additional seeds, and validation against ground-truth
network data, where available, are the most natural next steps.

\section{Conclusion}
\label{sec:conclusion}

Across 192 verified networks, the four prompting architectures formalized in
Eqs.~\eqref{eq:seq}--\eqref{eq:iter} produce structurally distinct conditional distributions over
edge sets, and their empirical signatures align with the theoretical distinctions between
factorizing and non-factorizing methods. Cultural framing alters network cohesiveness; political
affiliation dominates tie formation under sequential, local, and iterative generation while
global generation substitutes age; model scale induces a stable divergence ranking with the
smallest variant qualitatively distinct; and prompt language reshapes religion homophily under
Hindi while leaving political homophily nearly invariant. LLM-based social network generation
matches real graphs on clustering and modularity but encodes inbreeding homophily exceeding
empirical levels, and its outputs must therefore be interpreted in light of the formal
architecture, prompt language, and model scale that produced them.

{
\scriptsize
\setlength{\bibsep}{1pt}
\bibliographystyle{abbrvnat}
\bibliography{references}
}
\appendix
\newpage
\section{Persona Roster}
\label{app:roster}

The roster comprises 50 personas, each a structured profile $v_i \in \mathcal{A}$ as defined in
Section~\ref{sec:method}. Demographic marginals were sampled from U.S.\ adult population estimates
\citep{uscensus2023}; persona interests were synthesized using GPT-4o conditional on the other
five attributes. The roster is generated once and used unchanged across every experimental
condition, which makes it a controlled constant rather than an experimental variable. The
composition contains a slight Republican plurality ($n=26$ versus $n=24$ Democrats), with
Protestants ($n=19$) and the unreligious ($n=20$) as the two largest religious groups. The full
JSON roster, together with per-attribute marginals and the corresponding random baselines
$p^{\mathrm{base}}_a$ used in Eq.~\eqref{eq:samegroup}, is released in the supplementary material.

\section{Prompt Templates}
\label{app:prompts}

The four prompt templates $\pi_M$ for $M \in \{\mathrm{seq}, \mathrm{glob}, \mathrm{loc},
\mathrm{iter}\}$ are reproduced verbatim in the supplementary material, in all four languages used
in the language study (English, Spanish, Hindi, Japanese). Prompts were translated by native
speakers and back-translated for verification. The local-method neighborhood function $\nu$ uses a
fixed neighborhood size $k = 12$, sampled at random from $\mathcal{V} \setminus \{v_i\}$ with a
fixed RNG seed. The iterative method uses $T = 3$ rounds initialized from the sequential output.

\section{Reproducibility}
\label{app:repro}

The full experimental pipeline is implemented in Python and is reproducible from the command line
using three orchestration scripts: \texttt{run\_cultural\_study.py},
\texttt{run\_method\_study.py}, and \texttt{run\_language\_study.py}. The codebase is adapted from
the open-source repository accompanying \citet{chang2024llms} under its original license. All
generated networks are saved as adjacency lists with filenames encoding the full experimental
condition (\texttt{METHOD\_MODEL\_CONDITION\_SEED.adj}), making provenance unambiguous. Network
analysis is performed in NetworkX. All API calls use a fixed temperature of $\tau = 0.8$. Code and
the 192 verified adjacency lists are released as supplementary material. Two seeds per condition
yield $4 \times 1 \times 1 \times 2 \times 3 = 24$ cultural-study networks,
$4 \times 1 \times 3 \times 2 \times 3 = 72$ method-study networks, and
$4 \times 1 \times 4 \times 2 \times 3 = 96$ language-study networks; all 192 networks passed
verification as valid directed graphs over the fixed 50-node persona roster.

\section{Bias Analysis (Supplementary)}
\label{app:bias}

When the \texttt{--include\_reason} flag is enabled, the model produces a short natural-language
justification alongside each tie nomination. We analyze these reasons using the Bayesian
fighting-words method \citep{monroe2008fightin}, which fits a Dirichlet--multinomial model to two
text corpora and computes per-$n$-gram $z$-scores reflecting the strength of association with one
corpus over the other. This analysis is supplementary to the headline graph-metric findings
reported in the main text and is included to support interpretation rather than to derive new
claims.



\end{document}